\documentclass[sigplan, 10pt]{acmart}
\acmSubmissionID{<751>}
\AtBeginDocument{%
  }

\usepackage{graphicx}
\usepackage{subfig}
\usepackage{enumitem}
\usepackage{comment}
\usepackage{booktabs, amsmath, siunitx}
\usepackage[nameinlink,capitalize]{cleveref}
\usepackage{todonotes}
\usepackage{makecell}

\newcommand{\minisec}[1]{\noindent\textbf{#1.}}

\newcommand{\edit}[1]{#1}
\newcommand{\move}[1]{#1}
\newtheorem{theorem}{Theorem}
\title{Towards Efficient Flash Caches with Emerging NVMe Flexible Data Placement SSDs}

\author{Michael Allison, Arun George, Javier Gonzalez, Dan Helmick, Vikash Kumar, Roshan R Nair, \\Vivek Shah}

\affiliation{
\institution{Samsung Electronics}
\city{}
\country{}
}

\authornote{All author names are listed in alphabetical order.}

\renewcommand{\shortauthors}{Shah et al.}



\begin{document}
\acmYear{2025}\copyrightyear{2025}
\setcopyright{rightsretained}
\acmConference[EuroSys '25]{Twentieth European Conference on Computer Systems}{March 30--April 3, 2025}{Rotterdam, Netherlands}
\acmBooktitle{Twentieth European Conference on Computer Systems (EuroSys '25), March 30--April 3, 2025, Rotterdam, Netherlands}
\acmDOI{10.1145/3689031.3696091}
\acmISBN{979-8-4007-1196-1/25/03}
\begin{abstract}
NVMe Flash-based SSDs are widely deployed in data centers to cache working sets of large-scale web services. As data centers face increasing sustainability demands, such as reduced carbon emissions, efficient management of Flash overprovisioning and endurance has become crucial. Our analysis demonstrates that mixing data with different lifetimes on Flash blocks results in high device garbage collection costs, which either reduce device lifetime or necessitate host overprovisioning. Targeted data placement on Flash to minimize data intermixing and thus device write amplification shows promise for addressing this issue.

The NVMe Flexible Data Placement (FDP) proposal is a newly ratified technical proposal aimed at addressing data placement needs while reducing the software engineering costs associated with past storage interfaces, such as ZNS and Open-Channel SSDs. In this study, we explore the feasibility, benefits, and limitations of leveraging NVMe FDP primitives for data placement on Flash media in CacheLib, a popular open-source Flash cache widely deployed and used in Meta's software ecosystem as a caching building block. We demonstrate that targeted data placement in CacheLib using NVMe FDP SSDs helps reduce device write amplification, embodied carbon emissions, and power consumption with almost no overhead to other metrics. Using multiple production traces and their configurations from Meta and Twitter, we show that an ideal device write amplification of \textasciitilde1 can be achieved with FDP, leading to improved SSD utilization and sustainable Flash cache deployments.
\end{abstract}

\maketitle


\section{Introduction}
Caching is an intuitive and pervasive technique employed in modern large-scale web service architectures for high performance and better resource utilization. These web services range across various domains e.g., social networks, microblogging platforms, and emerging IoT applications, to name a few~\cite{berg2020cachelib, twemcache, mcallister2021kangaroo, BronsonACCDDFGKLMPPSV13:tao-fb}. \textit{The challenges in the design of the caching solutions for this domain are (1) the large working set size and (2) the problem of caching objects of different sizes, especially dominated by numerous small-sized objects~\cite{berg2020cachelib, mcallister2021kangaroo, twemcache}}. The use of Flash-based SSDs has become popular in the design of these caches given their excellent performance cost tradeoff compared to DRAM and HDD~\cite{TangHLKL15:ripq, EisenmanCPHSAK19:flashield, WongWMGLKSBBG24:baleen, berg2020cachelib, mcallister2021kangaroo}. However, managing the limited write endurance of Flash in these caches remains a challenge~\cite{LeeSHC15:f2fs, HeKAA17:ssd-contract, JungK13:ssd-expectations}. To maximize Flash lifetime in Flash caches, extensive research has gone into admission policies, application write amplification, and caching algorithms. However, the problem of device-level write amplification (DLWA) in Flash caches has not received much attention.

The problem of device-level write amplification is important today given the increased focus on data center carbon emissions. As multiple data center operators e.g., Amazon~\cite{amazon-sustainability}, Google~\cite{google-sustainability}, Meta~\cite{meta-sustainability}, and Microsoft~\cite{microsoft-sustainability} are aiming to achieve Net Zero greenhouse emissions, they are focusing on cutting down on embodied carbon emissions. Reductions in embodied carbon emission will constitute the bulk of data center emissions post their switch to renewable sources of energy~\cite{GuptaEHWL0W22:act}. Since Flash is more carbon-efficient than DRAM per bit~\cite{GuptaEHWL0W22:act, embodiedcarbon}, it is important to cache more data in Flash than DRAM and increase the lifetime of Flash devices for carbon-efficient deployment of Flash caches at scale. 

State of the art Flash caches~\cite{berg2020cachelib, mcallister2021kangaroo} employ specialized engines for caching small and large objects. A set-associative cache design is used to minimize the tracking overhead of numerous small objects while a log-structured design is used to cache large objects to generate Flash-friendly writes. These two cache designs have distinct write patterns on the SSD. The set-associative cache produces frequent updates in a random fashion while the log-structured cache produces infrequent updates in a sequential fashion on the SSD. To counteract high DLWA, production deployments of these caches underutilize the Flash device leading to a large embodied carbon footprint. Production deployments of CacheLib~\cite{berg2020cachelib} which is an open-source Flash cache used as a caching building block to build and deploy 100s of services at Meta only utilizes 50\% of the Flash capacity~\cite{berg2020cachelib,mcallister2021kangaroo}. \textit{Our analysis shows that the cause of high DLWA in these cache designs is the intermixing of data from the two different caching engines. Targeted data placement on Flash to isolate data from the specialized engines holds promise to reduce DLWA and embodied carbon footprint.}

The ratified NVMe Flexible Data Placement technical proposal~\cite{fdp_tp} is the latest technical proposal on data placement that incorporates lessons learned from previous proposals (e.g., Multi-Stream SSDs~\cite{kang2014multi}, Open-Channel~\cite{bjorling2017lightnvm}, ZNS~\cite{bjorlingAHRMGA21}) to improve adoption. It provides abstractions for isolating data on NAND media without incurring the software engineering cost of garbage collection or NAND media management. FDP is backward compatible so an application can run unchanged on it. This is particularly important for the adoption of FDP SSDs by production systems that favour stability and are sensitive to maintenance burden of emerging storage interfaces over time.

We designed and implemented data isolation modules by harnessing FDP features to reduce data intermixing in Flash media from Flash caches. \textit{Our key insight is that the high invalidation rate of set-associative cache design over a small LBA space can be harnessed along with device overprovisioning to ensure the availability of spare blocks for writing new incoming data for it.} This is important in maintaining a low predictable DLWA and reduces embodied carbon emissions. 

Our design utilizing FDP data placement features is non-invasive to the architecture of Flash caches. We observe that the specialized architecture of Flash caches, designed for both small and large objects, can efficiently tag the objects stored within it along the I/O path. This aligns well with the feature set of FDP which allows applications to experiment with data placement only.  Our design allows the automatic discovery of FDP features and adaptability to the SSD topology. It also enables the pluggability of various placement decisions to allow extensibility. 

We designed and implemented our data isolation features for small and large objects in a state-of-the-art open-source caching library CacheLib~\cite{berg2020cachelib}. \textit{Our changes have been merged in the upstream repository and deployed at scale ~\cite{fdp-pr}}. To quantify the gains of data isolation in Flash caches, we also devised a theoretical model for DLWA. We present a comprehensive evaluation of our design and implementation using multiple publicly available production workloads from Meta and Twitter, used in past research~\cite{berg2020cachelib, mcallister2021kangaroo, twemcache}, to quantify the benefits and limitations of our approach. Our experiments demonstrate that data separation in flash caches can result in a 2x reduction in SSD device costs and a 4x reduction in embodied carbon footprint. Moreover, our results also highlight opportunities to reduce the DRAM sizes in Flash cache deployments and explore multi-tenant deployments that were not possible earlier due to host overprovisioning.  At scale, this translates to massive cost benefits and embodied carbon emission reductions. 

Concretely, this paper makes the following contributions.
We review the concepts of the FDP storage interface, its limitations, and its connection to previous data placement proposals (Section \ref{sec:fdp}). We analyze the advantages of data segregation by leveraging FDP features in Flash cache architectures equipped with specialized engines for storing objects of varying sizes (Section \ref{sec:observations}). Additionally, we present a theoretical model to quantify DLWA. We present the design and implementation details of Flash cache data segregation by incorporating FDP features into CacheLib, a popular state-of-the-art open-source caching library, without altering its architecture or user-facing API (Section \ref{sec:implementation}). Our experiments show that separating small, hot data from large, cold data in Flash caches can reach an optimal DLWA of \textasciitilde1 without requiring any host overprovisioning, even with multiple challenging read- and write-intensive workloads with billions of small object accesses.

\section{Background}
\label{sec:background}
In this section, we highlight important concepts to facilitate a better understanding of the rest of the paper. We provide a summary of how SSDs work, the challenges associated with Flash caches, and the architecture of CacheLib~\cite{berg2020cachelib}.
\move{
\subsection{SSDs and Write Amplification}
\minisec{SSD Basics} A SSD NAND package is organised into dies, planes, blocks, and pages ~\cite{ssd_agarwal}. NAND SSDs cannot be directly overwritten due to the erase before write property. The erase operation happens in terms of erase blocks (EBs) (tens to hundreds of MBs) while the writes happen in terms of pages (16KB, 48KB, 64 KB, etc.). The Flash Translation Layer (FTL) in SSDs handles the overwrites by (1) writing (programming) the new data in a free page, (2)  invalidating the old data, and (3) updating the metadata to point to the new data. The metadata translates the logical addresses to physical addresses in the NAND media. Writing to logical addresses in an SSD creates invalid pages, that have to be reclaimed by a process called Garbage Collection (GC).

\minisec{SSD Garbage Collection (GC)} The garbage collection process is triggered whenever there is a scarcity of free blocks in the SSD. This process reads the remaining valid pages from an erase block and programs them to a new location. The now fully invalid erase block is available in the free pool. Any erase block in the free pool may be erased and programmed with the next incoming data to be written.  Garbage collection is an expensive operation and the energy consumed by the SSD is directly proportional to the number and duration of garbage collection operations.

\minisec{Device-level Write Amplification (DLWA)} DLWA is a metric used to quantify the amount of data that was written internally in the SSD compared to the actual amount of data sent by the host to the SSD. It can be calculated as follows:
\begin{equation}
    \text{DLWA} = \frac{\text{Total NAND Writes}}{\text{Total SSD Writes}}
\end{equation}

\minisec{Application-level Write Amplification (ALWA)} ALWA is a metric used to quantify the amount of data that was sent to the SSD to be written compared to the actual amount of data received to be written by the application. It can be calculated as follows:
\begin{equation}
    \text{ALWA} = \frac{\text{Total SSD Writes}}{\text{Total Application Writes}}
\end{equation}

\minisec{Importance of DLWA} The additional reads and writes from garbage collection interfere with the processing of other commands in the SSD affecting the QoS. Moreover, the additional NAND activity will consume the limited endurance of NAND media. A DLWA of 2 implies that for every 4 KB of data that the user writes, the FTL has written an extra 4 KB due to garbage collection. Since NAND media has a fixed number of Program and Erase cycles (P/E cycles) after which it can either only be read or becomes faulty, a DLWA of 2 causes the device's lifetime to be halved. Device-level write amplification impacts other SSD performance metrics, such as QoS, bandwidth, lifetime, reliability, and power consumption. It is often used as a simple proxy metric for monitoring SSD performance.
}
\subsection{DLWA and Carbon Emissions}
The lifetime of an SSD is inversely proportional to the device-level write amplification ~\cite{tbw_liftime}. A DLWA of 2 causes the SSD to fail twice as fast compared to DLWA of 1. A high DLWA results in premature SSD failure and requires frequent replacement of the device. \textit{SSD manufacturing produces millions of metric tonnes of CO2 emissions per year ~\cite{embodiedcarbon}. These emissions are broadly categorized as embodied carbon emissions.} With systems moving away from HDDs to SSDs, the need to reduce DLWA is crucial because the embodied carbon cost of SSDs is at least an order of magnitude larger than HDDs. Reduction of DLWA amortizes both capital costs and embodied carbon emissions of Flash-based systems at scale. 

High DLWA results from increased garbage collection operations to move valid pages to free up SSD blocks. Consequently, the SSD spends more time in the active state than in the idle state which results in a larger energy consumption~\cite{ssdpower, ssdenergy, sustainability-sdc}. Lower DLWA results in lower consumption of operational energy and translates to higher operational carbon efficiency. Although operational carbon efficiency optimization is important, big carbon efficiency gains are not expected from it. This is because SSDs are designed to be energy efficient and optimized to switch to idle state when not in use.

\subsection{Flash Caches and CacheLib}
\label{sec:background:flash-cache-cachelib}
\begin{figure}[!t]
    \centering
    \includegraphics[width=0.8 \linewidth]{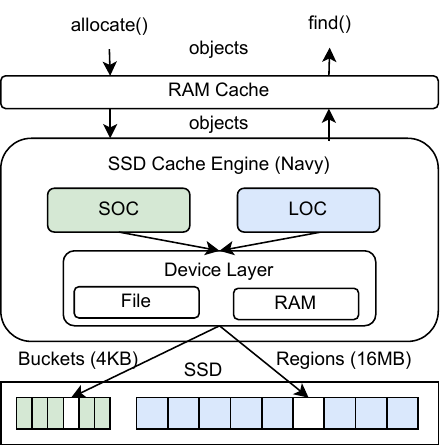}
    \caption{CacheLib Architecture Overview}
    \label{fig:cachelib}
\end{figure}

\minisec{Caching in Flash} Caching is widely employed in large-scale web services to provide high performance and reduce operational costs. Flash-based SSDs have become popular for caching large working sets of popular web services~\cite{berg2020cachelib, EisenmanCPHSAK19:flashield, TangHLKL15:ripq, WongWMGLKSBBG24:baleen} due to their excellent price performance tradeoff compared to DRAM and HDD. Caching on Flash is write-intensive since evictions upon read from DRAM translate to writes on Flash. Writes to Flash caches increase with the size of the working set, churn in keys, and reduction of DRAM sizes in the deployment. However, the limited write endurance of Flash coupled with unpredictable workloads pose a challenge in the design of these caches. 

Recent research has investigated the design of caching algorithms, admission policies, and application-level write amplification to manage the limited device endurance of Flash while delivering high hit ratios but device-level write amplification (DLWA) has been an understudied problem in this area. \textit{As sustainability challenges mount, Flash-based SSDs will become increasingly attractive compared to DRAM to cache large working set sizes at acceptable performance. However, before claiming Flash-based SSDs to be a panacea for sustainability, DLWA of Flash caches needs to be studied}.

\minisec{CacheLib Architecture} \textit{CacheLib~\cite{berg2020cachelib} is an open-source caching library that is widely used and deployed as a fundamental caching building block by 100s of services at Meta}. It employs a hybrid cache architecture (see Figure ~\ref{fig:cachelib}) to leverage both DRAM and Flash-based SSD to cache data hierarchically. DRAM is used to cache the most popular items while the SSD caches data that is less popular and evicted from the DRAM cache. The SSD cache is further decomposed into a small object cache (SOC) and a large object cache (LOC) to store objects of different sizes. The threshold for characterizing objects as small or large is configurable at deployment, along with the sizes of the DRAM and SSD. A single instance of CacheLib can consist of multiple DRAM and SSD cache engines, each with their configured resource budgets.

The SOC employs a set-associative cache design to perform in-place SSD writes in terms of buckets (typically aligned with 4 KB page size) and utilizes a uniform hashing function to minimize the overhead of tracking numerous small objects. The LOC employs a log-structured design to perform SSD-friendly writes in terms of large regions (16 MB, 256 MB, etc.) that align with erase block sizes. The LOC can be configured to use FIFO or LRU eviction policies. The strengths and weaknesses of the SOC and LOC complement each other. The LOC has SSD-friendly write patterns but has DRAM overheads for tracking objects, while the SOC has SSD-unfriendly write patterns and almost no overhead for tracking objects.\\  

\minisec{Challenges in Production Flash Caches and CacheLib}
The central challenge of Flash caches deployed in the wild is to manage the limited endurance of Flash while ensuring a high hit ratio and low indexing overhead. They have to deal with mixed workloads with objects of varying sizes and access patterns. \textbf{Large caching services typically handle billions of frequently accessed small items and millions of infrequently accessed large items ~\cite{berg2020cachelib, mcallister2021kangaroo, twemcache}}. The use of host overprovisioning and threshold admission policy is common for reducing DLWA. \textbf{In production CacheLib deployments, 50\% of the Flash capacity is overprovisioned to keep DLWA within acceptable levels of \textasciitilde1.3}~\cite{mcallister2021kangaroo, berg2020cachelib}. Increasing utilization of Flash with low DLWA is crucial for sustainable deployments of Flash caches at scale in the future.

\vspace{-2ex}
\section{NVMe Flexible Data Placement (FDP)}
\label{sec:fdp}
\subsection{Overview}
The ratified NVMe Flexible Data Placement technical proposal~\cite{fdp_tp} represents an evolution in the space of SSD data placement based on lessons learned in the wild over the past decade. It is a merger of Google's SmartFTL~\cite{smartftl} and Meta's Direct Placement Mode proposals to enable data placement on Flash media without the high software engineering costs of explicit garbage collection of ZNS~\cite{bjorlingAHRMGA21} and low-level media control of Open-Channel SSD proposals~\cite{bjorling2017lightnvm}. It borrows elements from the multi-streamed SSD interface~\cite{kang2014multi} that was proposed a decade ago but did not really take off due to a lack of industry and academic interest. It has been designed with backward compatibility in mind so that applications can work unchanged with it. The choice of leveraging data placement and evaluating its costs and benefits has been left to the application. This enables investment of engineering effort in a pay-as-you-go fashion instead of an upfront cost.

\subsection{Physical Isolation in SSDs with FDP}
\subsubsection{FDP Architectural Concepts}
\begin{figure}[!t]
  \centering
  \includegraphics[width=1.3 \linewidth]{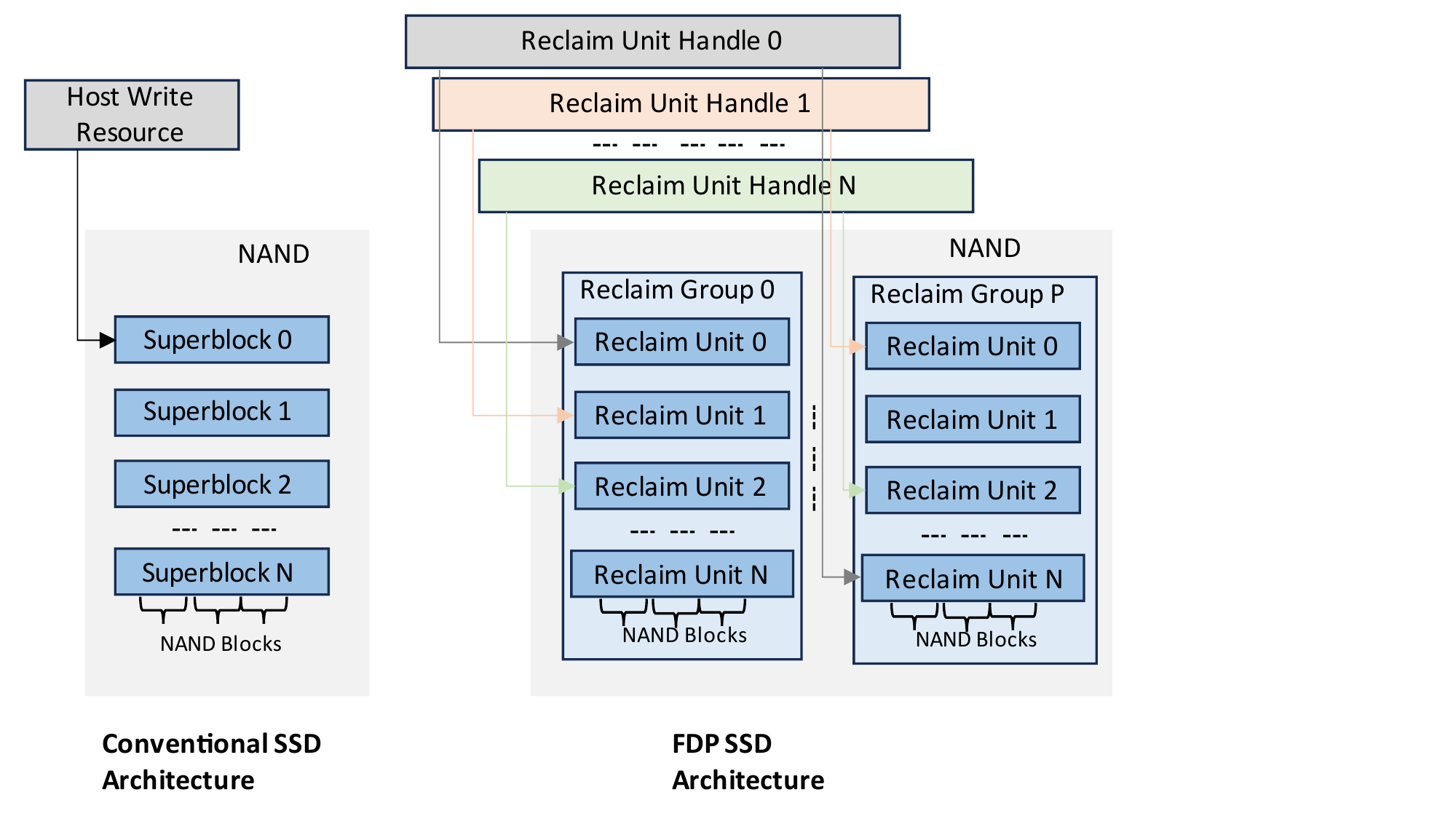}
  \caption{Conventional SSD vs FDP SSD Architecture.} 
  \vspace{-2ex}
  \label{fig:fdp-arch}
\end{figure}

The Flexible Data Placement interface provides abstractions to the host to group data on the device with a similar expected lifetime (e.g., death time). The interface introduces the following concepts to expose the SSD physical architecture (see Figure \ref{fig:fdp-arch}).

\minisec{Reclaim Unit (RU)} The NAND media is organized into a set of reclaim units where a reclaim unit consists of a set of blocks that can be written. A reclaim unit will typically consist of one or more erase blocks but no guarantees are made in the proposal towards this end. The size of an RU is decided by the SSD manufacturer. In this paper, our device has superblock-sized RUs where a superblock is a collection of erase blocks across the planes of dies in the SSD. If an SSD has 8 dies each with 2 planes and 2 erase blocks per plane, the superblock will consist of 32 erase blocks.

\minisec{Reclaim Group (RG)} A reclaim group is a set of reclaim units.

\minisec{Reclaim Unit Handles (RUH)} A reclaim unit handle is an abstraction in the device controller similar to a pointer that allows host software to point to the reclaim units in the device. Since a reclaim unit is not directly addressable by the host, the host software uses the reclaim unit handles to logically isolate data. The device manages the mapping of reclaim unit handles to a reclaim unit and has complete control over this mapping. The number of RUHs in the device determines the number of different logical locations in the NAND where the host software can concurrently place data.

\minisec{RUH Types} \move{
The FDP interface specifies two types of reclaim unit handles, each offering distinct data movement guarantees during garbage collection, along with their respective tradeoffs. During garbage collection, the RUH type is used to determine the source and destination RUs of data to be moved. FDP defines two RUH types namely,
\begin{enumerate}[leftmargin=*]
    \item \textbf{Initially Isolated} - All the reclaim units within a reclaim group pointed to by the RUHs of this type are candidates for data movement. For multiple RUHs of initially isolated type, data starts off being isolated from data written using another RUH of initially isolated type. However, upon garbage collection valid data written using these two handles can be intermixed. This type is the cheapest to implement on the SSD controller as it does not require explicit tracking of data written using RUHs and imposes the least constraints on data movement during garbage collection. 
    \item \textbf{Persistently Isolated} - All the reclaim units within a reclaim group that have been written utilizing the RUH are the only candidates for data movement upon garbage collection. This RUH type provides a stronger guarantee of data isolation but is expensive to implement on the controller as it requires explicit tracking of data written using RUHs and imposes more constraints on data movement during garbage collection.
\end{enumerate}
\minisec{Example} Consider a write pattern using two RUHs, RUH0 and RUH1 where RUH0 has written LBAs to RU0 and RU1 while RUH1 has written LBAs to RU2. For simplicity, let us assume that all the RUs belong to the same reclaim group. If RUH0 and RUH1 are of initially isolated type, then upon garbage collection valid data from RU0, RU1 and RU2 are candidates for movement and can be intermixed. If RUH0 and RUH1 are of persistently isolated type, then only data from RU0 and RU1 can be intermixed upon garbage collection while the data in RU2 is isolated from data in RU0 and RU1.
}

\minisec{FDP Configurations}  An FDP configuration defines the RUHs, RUH type (Initially or Persistently Isolated), their association to RGs, and the RU size. \textit{The FDP configurations available on the device are predetermined by the manufacturer} and cannot be changed. This paper uses an SSD with a single FDP configuration of 8 Initially Isolated RUHs, 1RG and RU size of 6GB. A device can support multiple configurations that can be chosen by the host.

\subsubsection{Data Placement with RUHs}
In this section, we highlight important aspects of the FDP interface that influence data placement designs by the host.\\
\minisec{Physically Isolating Logical Blocks with RUHs}
The FDP storage interface does not introduce any new command sets to write to the device. Instead, a new data placement directive has been defined that allows each write command to specify a RUH. Thus, the host software can use the RUH to place a logical block in a RU utilizing the RUH. By allowing the host to dynamically associate a logical block with a RU, FDP enables flexible grouping of data based on varying temperature and death time (e.g., hot and cold data separation) or different data streams (e.g., large streams and small journals). This facilitates writing to different RUs in a physically isolated manner. By careful deallocation of all the data in a previously written RU, the host can achieve a DLWA of \textasciitilde1.

During namespace creation, the host software selects a list of RUHs that are accessible by the created namespace. Since FDP is backward compatible, a default RUH is chosen by the device for a namespace if it is not specified. Data is placed in this RUH in the absence of the placement directive from the host. Read operations remain unchanged as before. Writes in FDP can cross RU boundaries. If a write operation overfills an RU because the RU is written to its capacity, the device chooses a new RU and updates the mapping of the RUH to the new RU. Although this process is not visible to the host, the event is logged by the SSD in the device logs that the host can examine. 

\minisec{Managing Invalidations and Tracking RUs}
Since FDP does not focus on garbage collection but purely data placement, it does not introduce any new abstractions for erase operations. As in conventional SSDs, LBAs are invalidated or dealloacted in two ways, (1) by overwriting an LBA, (2) by explicitly using a trim operation over one or many LBAs. If all the data in a RU is invalidated, then the RU is erased for future writes and no logical blocks have to copied across RUs upon garbage collection. Since the host software can only access RUHs and not an RU, in order to perform fine-grained and targeted deallocation of RUs, the host software needs to track the LBAs that have been written to an RU together and deallocate those. The FDP specification also allows the host to query the available space in an RU which is currently referenced by the RUH.

\subsection{FDP Events and Statistics} FDP provides an elaborate set of events and garbage collection statistics for the host to track the FDP related events in the SSD. These help the host to be aware of device-level exceptions and make sure that both host and device are in sync regarding data placement.

\subsection{FDP and Other Major Data Placement Proposals}
\edit{
NVMe FDP technical proposal was conceived based on lessons learnt from integrating software stacks with the past data placement proposals. It has been designed to focus on data placement to allow host software stack to perform data segregation while leaving NAND media management and garbage collection to the SSD controller. In Table~\ref{tab:fdp-data-placement-tech}, we outline some of the key differences between the major data placement proposals of the past years. More details can be found in some of the recent industry presentations on FDP~\cite{sdc-fdp-dan, sdc-fdp-mike}.
}

\begin{table*}[!ht]
    \centering
    \edit{
    \begin{tabular}{|p{0.15\textwidth}|p{0.18\textwidth}|p{0.18\textwidth}|p{0.15\textwidth}|p{0.18\textwidth}|}
        \hline
    Characteristic & \textbf {Streams~\cite{kang2014multi}} & \textbf {Open-Channel~\cite{bjorling2017lightnvm}} & \textbf{ZNS~\cite{zns, bjorlingAHRMGA21}} & \textbf{FDP~\cite{fdp_tp}} \\ \hline
    Supported write patterns & Random, Sequential & Random, Sequential & Sequential & Random, Sequential \\ \hline
    Data placement primitive & Using stream identifiers & Using logical to physical address mapping by host & Using zones & Using reclaim unit handles \\ \hline
    Control of garbage collection & SSD-based without feedback to host & Host-based & Host-based & SSD-based with feedback through logs \\ \hline
    NAND media management by host & No & Yes & No & No \\ \hline
    Can run applications unchanged & Yes & No & No & Yes \\ \bottomrule
    \end{tabular}
    \caption{High-Level Comparison of Major Data Placement Proposals.}
    \label{tab:fdp-data-placement-tech}
    }
    \vspace{-2ex}
\end{table*}

\edit{
\subsection{Limitations}
\begin{enumerate}[leftmargin=*]
\item {\textbf{New and evolving technology.}} The FDP technical proposal was ratified at the end of 2022, and some devices from Samsung, such as the PM9D3a~\cite{pm9d3a} are emerging on the market with support for it, along with offerings from other vendors. Due to the relatively recent ratification, the proposal may undergo modifications over time to include extensions for desirable features.
\item {\textbf{Lack of host control over garbage collection.}} FDP was designed specifically for data placement while allowing hosts to perform random writes to LBAs, enabling the SSD to manage garbage collection. Consequently, the host has no control over the garbage collection process in the SSD, aside from invalidating LBAs by deallocating or overwriting them. Note that this limitation only applies in scenarios where the host can achieve greater performance gains by managing garbage collection more efficiently than the SSD, rather than focusing solely on data placement.
\item {\textbf{Requires device overprovisioning and mapping table in SSD.}} As in conventional SSDs today, FDP SSDs will also require a mapping table in DRAM to support transparent mapping of logical to physical addresses. Moreover, NAND overprovisioning in the device is required for acceptable performance in the absence of host-based garbage collection. This is a limitation when the proposal is viewed from the lenses of the cost of fabrication of FDP SSDs.
\end{enumerate}
}

\section{Why FDP Matters for CacheLib and Hybrid Caches?}
\label{sec:observations}
In this section, we discuss the fit of FDP and the opportunities afforded by it for CacheLib and hybrid caches based on the analysis of CacheLib's Flash Cache architecture, web service caching deployments, and workloads.

\subsection{Insights and Observations}
\begin{figure}[!t]
  \centering
  \vspace{-3ex}
  \subfloat[a][SSD cross-section without FDP]{\includegraphics[width=0.9\linewidth]{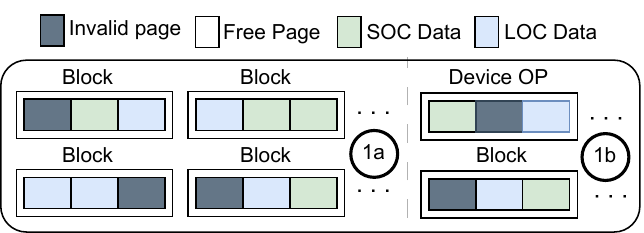} \label{fig:non-fdp-cross}} \hfill \\
  \vspace{-2ex}  
  \subfloat[b][SSD cross-section without FDP]{\includegraphics[width=0.9\linewidth]{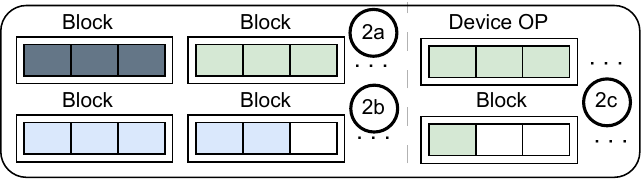} \label{fig:fdp-cross}} \hfill \\   
  \caption{SSD cross-section. {\Large \textcircled{\small 1a}} shows the intermixing of LOC’s sequential and cold data with SOC’s random and hot data in SSD blocks. {\Large \textcircled{\small 1b}} shows the inefficient use of device OP by both LOC and SOC data. {\Large \textcircled{\small 2a}} shows that with SOC data being segregated, invalidation of its data can result in free SSD blocks.  {\Large \textcircled{\small 2b}} shows that with FDP, LOC data which is written sequentially will not cause DLWA. {\Large \textcircled{\small 2c}} shows the efficient use of device OP exclusively by SOC data to cushion SOC DLWA.} 
  \label{fig:ssd-cross}
  \vspace{-2ex}
\end{figure}

\minisec{\textit{Insight 1: Intermixing of SOC and LOC data leads to high DLWA}}
Large cache items are written into the LOC in a log-structured fashion utilizing a FIFO or LRU eviction policy. This results in a sequential write pattern to the SSD. Small cache items are written into SOC buckets using a uniform hash function. Each item insert causes the entire SOC bucket (size is configurable but default is 4 KB) to be written to the SSD. Contrary to LOC, SOC writes generate a random write pattern to the SSD. 

For workloads with large working set sizes and key churn, the Flash cache layer receives writes due to evictions from the RAM cache~\cite{berg2020cachelib,mcallister2021kangaroo}. For workloads dominant in small object accesses, this segregation leads to an infrequent and cold data access pattern in the LOC together with a frequent and hot data access pattern in the SOC. This leads to the intermixing of LOC's sequential and cold data with SOC's random and hot data in a single SSD block (Figure \ref{fig:non-fdp-cross} {\Large \textcircled{\small 1a}}) causing high DLWA upon garbage collection. \\

\minisec{\textit{Insight 2: The use of host overprovisioning as a control measure for DLWA is inefficient}}
As explained in Section \ref{sec:background:flash-cache-cachelib}, \textit{CacheLib deployments utilize a host overprovisioning of almost 50\% of the SSD to limit DLWA to an acceptable value of $\sim$1.3}. This is inefficient from both cost and carbon efficiency perspectives. The LOC data due to its sequential and cold access pattern does not need any host or device overprovisioning for a DLWA of 1. Without host overprovisioning the only extra space available to help control DLWA is the device overprovisioning space. The random SOC data would benefit the most from the device overprovisioned space because it is small, hot and updated frequently. However, the intermixing of SOC and LOC results in an inefficient use of the device overprovisioning space (Figure \ref{fig:non-fdp-cross} {\Large \textcircled{\small 1b}}) as both the SOC and LOC data share it causing unnecessary data movement. \\

\minisec{\textit{Insight 3: High SOC invalidation and its small size can be harnessed to control DLWA}}
A smaller SOC size on devices leads to fewer buckets and a higher rate of key collisions. Since the entire SOC bucket of 4KB is written out, a larger SOC bucket invalidation rate is SSD-friendly because it leads to more SSD page invalidation. If only invalidated SOC data resided in an SSD erase block, this would result in the entire erase block freeing itself up and not needing movement of valid data. For workloads dominant in small object accesses, a high invalidation of SOC happens but the SOC data in erase blocks is intermixed with LOC data. This prevents the SSD from taking advantage of the updates occurring in the SOC buckets over a small LBA space. \\

\minisec{\textit{Insight 4: Data placement using FDP can help CacheLib control DLWA}}
FDP can be utilized by CacheLib to separate the SOC and LOC data in the SSD using different reclaim unit handles. This allows the LOC data and SOC data to reside in mutually exclusive SSD blocks (reclaim units). Such a design will have the following benefits,
\begin{enumerate}[noitemsep, topsep=1pt, partopsep=0pt,leftmargin=*]
    \item The SSD blocks containing LOC data get overwritten sequentially resulting in minimal data movement and DLWA (Figure \ref{fig:fdp-cross} {\Large \textcircled{\small 2b}}). If LOC data resides in separate reclaim units than SOC data, the device overprovisioning space can be used exclusively by SOC data.
    \item The ideal behaviour of SOC data invalidating only itself (Insight 3) can be realized by segregating it into separate reclaim units (Figure \ref{fig:fdp-cross} {\Large \textcircled{\small 2a}}). A smaller SOC size leads to a greater invalidation rate causing most of the SOC data in the SSD erase block being invalid. This leads to minimal live data movement and DLWA. As the SOC size increases we expect an increase in DLWA even with LOC and SOC segregation across reclaim units.
    \item The ideal utilization of device overprovisioning space (Insight 2) is possible with FDP (Figure \ref{fig:fdp-cross} {\Large \textcircled{\small 2c}}). SOC data can use the overprovisioned space to cushion DLWA. When the SOC size is smaller than the device overprovisioning space we expect a DLWA of $\sim$1 since there is at least one spare block available for each block of SOC data.
    \item The separation of LOC and SOC data in the SSD does not necessitate a change in the CacheLib architecture and API. Therefore, we expect no change in the application-level write amplification (ALWA).
\end{enumerate}
\vspace{1ex}

\minisec{\textit{Insight 5: Initially Isolated FDP devices will suffice in controlling the DLWA in CacheLib}}
With the separation of LOC and SOC data within the SSD, the only live data movement will be due to SOC data. Irrespective of whether the SSD is initially isolated or persistently isolated only SOC data would reside in reclaim units used for garbage collection. Therefore, the isolation of LOC and SOC data would be preserved regardless. \\

\minisec{\textit{Insight 6: Data placement using FDP can help reduce carbon emissions in CacheLib}}
Embodied carbon emissions account for the major chunk of carbon emissions compared to operational carbon emissions. The DLWA gains from using FDP-enabled CacheLib leads to an improved device lifetime. This results in fewer device replacements during the system lifecycle leading to reduction in embodied carbon emissions.

Fewer garbage collection operations are the reason for the DLWA gains with FDP-enabled CacheLib. For a fixed number of host operations, fewer data migrations result in fewer total device operations. The reduction in total operations requires the device to spend fewer cycles in the active state leading to a lower SSD energy consumption and reduced operational carbon footprint~\cite{ssdenergy, ssdpower}.

\subsection{Theoretical Analysis of FDP-enabled CacheLib DLWA and Carbon Emissions}
\label{theorem:DLWA}
We formulate a theoretical model of DLWA and carbon emissions for SOC and LOC data segregation in CacheLib using the insights of the previous section. We assume the DLWA of LOC data is $\sim$1. Additionally, we use the fact that only SOC data will use the device overprovisioning space and item insertions to the SOC buckets follow a uniform hash function. To simplify our analysis, we assume that the uniform hash function used in CacheLib is fairly well-behaved. Modelling DLWA by estimating live data movement for a uniform random workload has been used proposed before~\cite{DayanBB15}. We extend that methodology to model the SOC DLWA that translates to the DLWA for FDP-enabled CacheLib as the LOC does not contribute to DLWA. The derivation of the theorems in this section is available in Appendix \ref{ref:appendix:theoretical-model}.
\begin{theorem}
    The DLWA for FDP-enabled CacheLib using SOC and LOC data segregation is,
    \vspace{-0.75em}
    \begin{equation*}
        \text{DLWA} = \frac{1}{1- \delta}
    \end{equation*}
    \vspace{3ex}
    
    \noindent where $\delta$ denotes the average live SOC bucket migration due to garbage collection and is given by,
    \begin{equation*}
        \delta = - \frac{\text{S}_{\text{SOC}}}{\text{S}_{\text{P-SOC}}} \times \mathcal{W} (- \frac{\text{S}_{\text{P-SOC}}}{\text{S}_{\text{SOC}}} \times e^{- \frac{\text{S}_{\text{P-SOC}}}{\text{S}_{\text{SOC}}}} )
    \end{equation*}
    where $\text{S}_{\text{SOC}}$ is the total SOC size in bytes,  $\text{S}_{\text{P-SOC}}$ is the total physical space available for SOC data including device overprovisioning in bytes and $\mathcal{W}$ denotes the Lambert W function.
\end{theorem}

\subsubsection{Modelling CO2 emissions (CO2e) for FDP-enabled CacheLib}
The total carbon footprint is the sum of embodied and operational carbon emissions.
    \begin{equation*}
        \text{Total}_{\text{CO2e}} = \text{C}_{\text{embodied}} + \text{C}_{\text{operational}}
    \end{equation*}
\label{theorem:embodied-co2e}
\begin{theorem}
    The embodied carbon emissions from using CacheLib by accounting for SSD replacement during the system lifecycle of T years and rated SSD warranty of $\text{L}_{\text{dev}}$ years is,

    \begin{equation*}
        \text{C}_{\text{embodied}} = \text{DLWA} \times \text{Device}_{\text{cap}} \times \frac{T}{\text{L}_{\text{dev}}} \times \text{C}_{\text{SSD}}
    \end{equation*}
    where $\text{Device}_{\text{cap}}$ is the physical capacity of the device. \\
    $\text{Host}_{\text{cap}} = \text{Device}_{\text{cap}} \times (1 - \text{Total}_{\text{op}})$ denotes the SSD capacity used by the host system in GB, $\text{H}_{\text{op}} \text{ and } \text{D}_{\text{op}} \in [0,1) $ is the fraction of host overprovisioning and device overprovisioning and $\text{C}_{\text{SSD}}$ is the amount of CO2e (Kg) per GB of SSD manufactured.
    
\end{theorem}

Operational Energy can be converted to CO2 emission (CO2e) using the greenhouse equivalence calculator ~\cite{greenhouse_calc}. The operational energy consumed can be modelled by estimating the time spent in idle and active states ~\cite{ssdenergy}. The time spent in active states is proportional to the total number of device operations during the period in question,

\begin{theorem}
Operational energy is proportional to the total number of garbage collection events.
\begin{equation*} 
    \mathcal{E}_{\text{operational}} \propto \mathcal{E}(\text{Host}_{\text{operations}}) + \mathcal{E}(\text{Device}_{\text{migrations}})
    \end{equation*}
    where, $\text{Device}_{\text{migrations}}$ is the number of garbage collection operations triggered in the SSD. 
\end{theorem}

\section{Design and Implementation}
\label{sec:implementation}
\edit{
In this section we outline the design principles, implementation details, and lessons learned while building FDP-aware SOC and LOC data segregation in CacheLib.

\subsection{Design Principles}
CacheLib is a popular building block for various caching services with diverse use cases. It is important to design a minimally invasive, adaptable, and maintainable solution for SOC and LOC data segregation. We applied the following design principles while building FDP-based data segregation support in CacheLib.

\minisec{\textit{1. Keep it simple~\cite{Lampson83:simple}}}
This guiding principle enabled us to merge our changes upstream given the diverse use cases of CacheLib and the need for stability and maintainability of the codebase.

\minisec{\textit{2. FDP is just another storage technology for CacheLib}}
The design should seamlessly support deployments and setups where FDP is not used, requiring minimal configuration changes to remain user-friendly. This ensures that CacheLib maintains backward compatibility for users who do not utilize FDP SSDs.

\minisec{\textit{3. Allow software extensibility for various data placement technologies}}
The design should be generic and extensible to allow existing and future modules in CacheLib to segregate data and  experiment with various data placement policies and decisions.

\minisec{\textit{4. Allow hardware extensibility for evolving data placement technologies}}
Since the FDP specification is new and expected to evolve over time, its use should be localized in the CacheLib architecture to minimize changes to the codebase over time as the hardware technology evolves.
}

\subsection{Placement Handles}
We introduce the abstract concept of \texttt{Placement Handle} to CacheLib's SSD I/O path to achieve FDP-based data placement while preserving backward compatibility. \texttt{Placement handles} allow various consuming modules to segregate data. The set of available placement handles are allocated from a data placement aware device layer upon initialization. Such an abstraction hides the semantics of the underlying data placement technology e.g., FDP providing hardware extensibility. If FDP is not enabled on the underlying device, a default placement handle is used to indicate no placement preference.

\subsection{Placement Handle Allocator}
\begin{figure}[!t]
    \centering
    \includegraphics[width=0.8\linewidth]{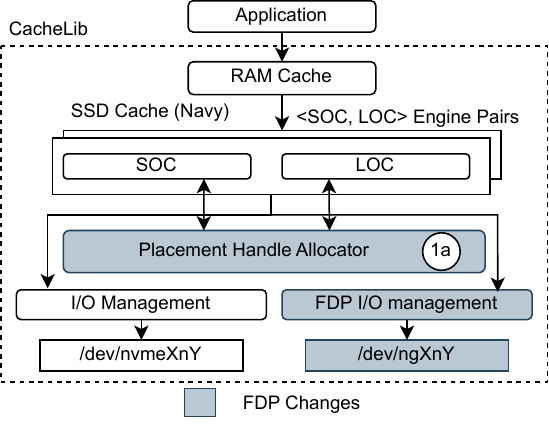}\hfill
    \caption{CacheLib I/O Path. {\Large \textcircled{\small 1a}} denotes the placement handle allocator that is responsible for allocating placement handles that consume FDP.}  
    \label{fig:cachelib-io-path}
\end{figure}
We implement a \texttt{Placement Handle Allocator} (Figure \ref{fig:cachelib-io-path} {\Large \textcircled{\small 1a}}) that is responsible for allocating placement handles to any module that wishes to use data placement. If FDP is enabled in CacheLib and the underlying SSD supports FDP, this module assigns an available <RUH, RG> pair, referred as a Placement Identifier (PID) in the FDP specification, to the placement handle it allocates. This abstracts out FDP semantics from the consumers of FDP. If the underlying SSD does not support FDP, the default handle will be allocated. This indicates that there is no placement preference. In CacheLib, SOC and LOC in each I/O engine pair get different allocation of placement handles during initialization. Modules that are minor consumers of SSD, e.g., the metadata, do not state their placement preferences, so the default reclaim unit handle is assigned to them. The introduction of the placement handle allocator provides software extensibility and flexibility.

\subsection{FDP Aware I/O Management}
The SOC and LOC instances tag their I/Os with unique placement handles. The FDP-aware device layer translates these handles to the corresponding FDP Placement Identifier (PID). The PIDs are further translated to the NVMe specification placement directive fields (DSPEC and DTYPE fields \cite{NVMeCS}), attached to the uring\_cmd \cite{uring-cmd} I/Os and then submitted. We use the I/O Passthru~\cite{io-passthru, JoshiG0KRGLA24:iopassthru} mechanism to send FDP-enabled I/Os to the Linux kernel. We use an io\_uring~\cite{axboe2019efficient} queue pair per worker thread so that I/Os can be sent to the kernel without synchronization or concurrency challenges in the submission and completion queues of io\_uring. The FDP aware I/O management layer provides hardware extensibility by abstracting the layout of an FDP-enabled SSD.

\subsection{Lessons Learned and Future Directions}
\edit{In this section, we outline a few lessons learnt over time as we integrated FDP features in CacheLib. Some of these lessons can be useful for future directions of work.\\

\minisec{\textit{1. FDP specialized LOC eviction policy in CacheLib did not provide much benefit}}
We explored the notion of a specialized FDP eviction policy for CacheLib by building reclaim unit size awareness in the region-based eviction policy of the LOC. The FDP specification provides the required semantics for the host to track writes to a reclaim unit. Utilizing it, we can track the LOC regions that belong to a reclaim unit and invalidate multiple regions in a reclaim unit together. This can be paired with a \texttt{TRIM} command to free an entire reclaim unit and aid the garbage collection process on the SSD. Early exploration of this policy showed minimal gains and was shelved. We speculate that such a policy could be beneficial in cases where reclaim units are smaller in size.}\\

\edit{
\minisec{\textit{2. Dynamic and adaptive data placement is outperformed by simple static solutions}}
Using FDP event logs, the host can inform itself of garbage collection operations in the SSD. This allows host software to understand the impact of its data placement decisions in real time. A host software stack can utilize the logs to build a feedback loop to understand its placement decisions and adapt accordingly. We explored some dynamic data placement policies using various load balancing and data temperature techniques early in the project. However, we saw minimal gains compared to the engineering complexity for such an effort over a static predefined placement handle for segregating SOC and LOC data for the small object dominant hybrid workloads.
}

\section{Evaluation}
\label{sec:evaluation}
In this section, we explore the benefits and limitations of segregating data in the small object cache and the large object cache of CacheLib on FDP SSDs. 
\subsection{Experimental Setup}
This section describes the setup of the system, the workloads for our experiments, and the metrics used in our evaluation. \\
\minisec{Hardware and Software Setup}
We use two different machines in our experimental setup with similar hardware characteristics. Both machines have two 24-core Intel Xeon Gold 6432 processors with \textasciitilde 528 GB of DRAM. Each machine uses a 1.88 TB Samsung PM9D3 SSD with a firmware version that supports the FDP specification. The FDP configuration on the device supports 2 namespaces, 1 RG and 8 initially isolated RU handles that can be used concurrently. For all experiments, we create a single namespace and map all the RU handles to it. Each RU is \textasciitilde 6 GB in size. One of the machines runs Ubuntu 22.04 with a 6.1.64 Linux kernel, while the other machine runs CentOS 9 with a 6.1.53 Linux kernel. We use \texttt{nvme-cli}~\cite{nvme-cli} version 2.7.1 for configuring FDP features on the SSD.\\
\minisec{System Comparisons}
For our experiments, we use the main branch of the CacheLib repository which contains our upstreamed FDP-based data placement changes to segregate SOC and LOC \cite{fdp-pr, cachelib-repo}. We use \texttt{nvme-cli} to enable and disable FDP features on the controller to contrast between a conventional SSD and an FDP SSD. In the rest of the paper, we use the following terms to highlight the system comparisons under test,
\begin{enumerate}[noitemsep, topsep=1pt, partopsep=0pt]
    \item \textbf{FDP:} FDP-based data segregation enabled in CacheLib and FDP configuration enabled on the SSD.
    \item \textbf{Non-FDP:} FDP-based data segregation disabled in CacheLib and FDP configuration disabled on the SSD.
\end{enumerate}
We use the CacheBench workload generator and trace replaying tool~\cite{cachebench} to run the workloads. CacheBench is an application that invokes the CacheLib cache API in the same process and can be used to run captured traces or generate benchmarks. \edit{All the scripts used to run experiments are available publicly~\cite{cachelib-devops}.}\\
\minisec{Metrics}
We focus on DLWA as the primary metric to evaluate the efficacy of our data segregation changes since it has a direct correlation to endurance and embodied carbon emissions. We measure DLWA by using the \texttt{nvme-cli} tool to query log pages (\texttt{nvme get-log}) from the SSD controller that tracks the host bytes written to it and the device bytes that were written on NAND media over an interval of 10 minutes. We reset the SSD to a clean state before every experiment by issuing a TRIM for the entire device size. We use the CacheBench tool to measure and report throughput, latency, DRAM and NVM cache hit ratios, and ALWA for our experiments.\\
\minisec{Workloads}
Our experiments use sampled 5 day anonymized traces from Meta's key-value cache (KV Cache) cluster \cite{berg2020cachelib, cachelib-traces} and 7-day anonymized traces from Twitter's cluster12 \cite{twemcache} that are publicly available for research and experimentation. \textbf{KV Cache} is a read-intensive workload where the GETs outnumber SETs by a 4:1 ratio. For KV Cache, we use \textasciitilde42 GB of DRAM and \textasciitilde930 GB of SSD (50\% device utilization) for caching as the default setup. Twitter's cluster12 workload is write-intensive, where the SETs outnumber GETs by a 4:1 ratio. For the \textbf{Twitter workload}, we use \textasciitilde16 GB of DRAM and \textasciitilde930 GB of SSD (50\% device utilization) as the default setup. To stress the SSDs more and generate high DLWA scenarios in a shorter duration, we generated an additional write-only KV cache workload by removing the GET operations from the KV cache trace so it almost exclusively consists of SET operations. We refer to it as the \textbf{WO KV Cache} workload and use the same DRAM and SSD configurations as mentioned for KV Cache. 

\subsection{FDP-based segregation achieves a DLWA of $\sim$1}
\label{sec:eval:kv-cache:waf-50}
\begin{figure}[!t]
  \centering
  \includegraphics[width=0.8\columnwidth]{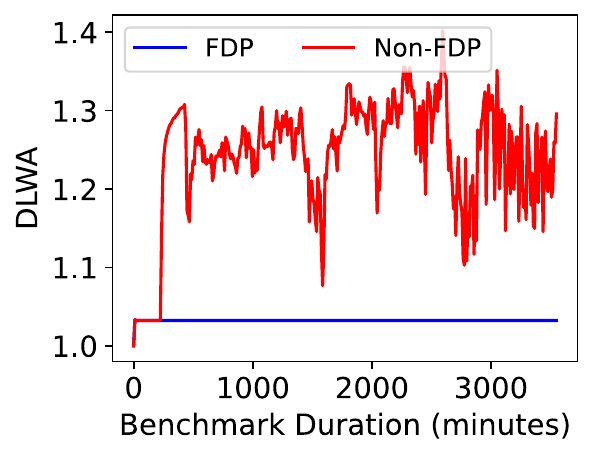}
  \caption{DLWA over 60 hours with the KV Cache workload using 50\% device utilization, 42GB of RAM and 4\% SOC size. FDP-based segregation results in a 1.3x reduction in DLWA.} 
  \label{fig:waf:kv-waf-50}
  \vspace{-3ex}
\end{figure}
We run the KV Cache workload with the default DRAM cache size of \textasciitilde42 GB and SSD cache size of \textasciitilde930 GB out of a total size of 1.88 TB SSD for an effective utilization of 50\%. The SOC size in the default configuration is set to 4\% of the SSD size. Figure \ref{fig:waf:kv-waf-50} shows the interval DLWA over a run duration of more than 2 days. We can see that the SOC and LOC data segregation into two different reclaim unit handles helps to lower DLWA from 1.3 observed without data segregation to 1.03. This validates our analysis that segregating the sequential and cold write pattern of LOC from the random and hot write pattern of SOC by utilizing device overprovisioning helps control DLWA. \textit{Thus, we achieve an ideal DLWA of $\sim$1 with the FDP-based segregation in CacheLib.}

\subsection{FDP-based segregation enables better SSD utilization without affecting performance}
\label{sec:eval:waf:kv-varying-util}
\begin{figure*}[!t]
        \begin{minipage}{0.9\linewidth}
        \hspace{-1em}
        \centerline{\includegraphics[width=0.95\linewidth]{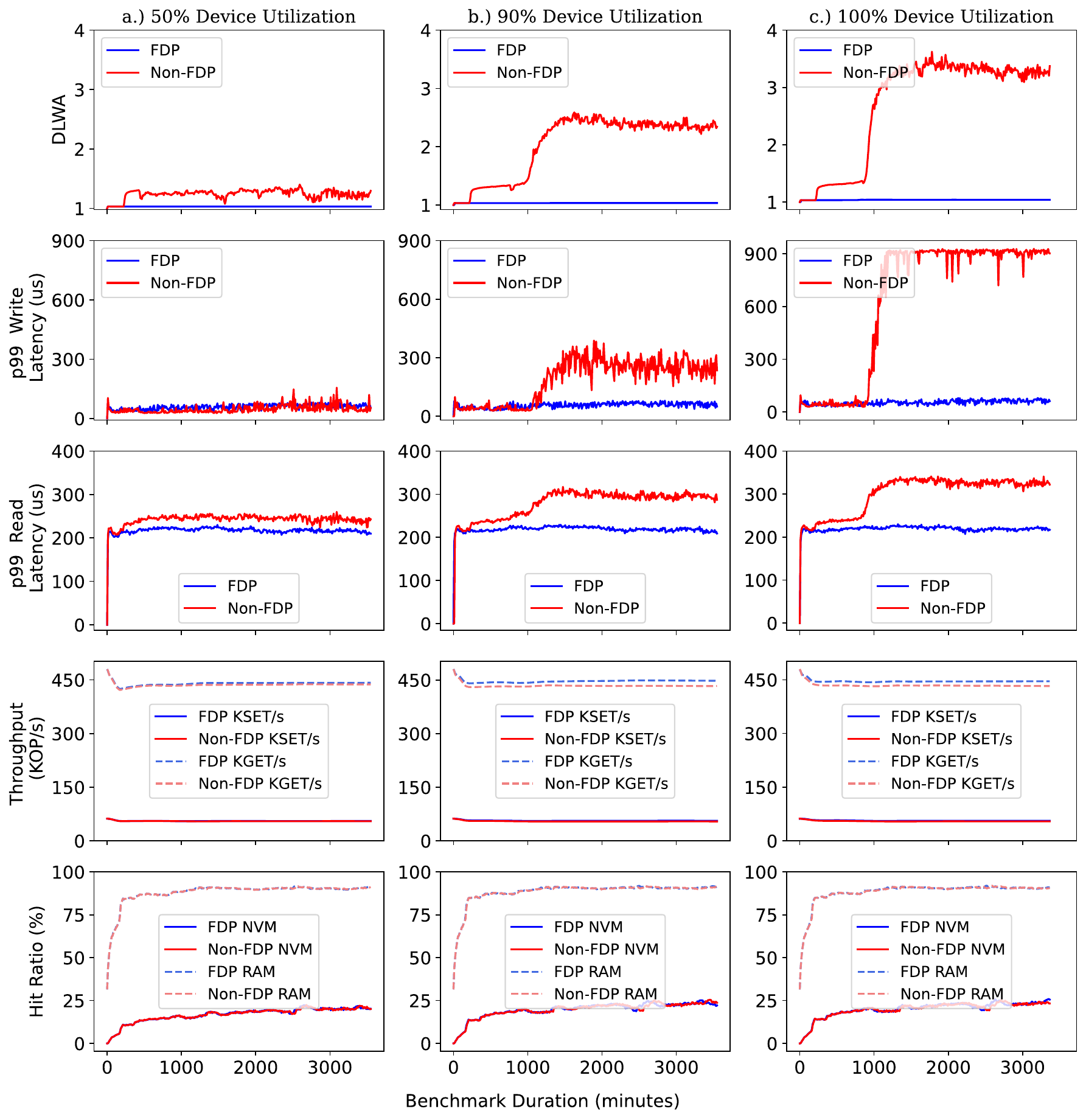}}
        \caption{Effect of varying SSD utilization for caching with KV Cache Workload on DLWA and other CacheLib performance metrics like throughput, p99 read and write latency, and DRAM and SSD cache hit ratios. FDP-based segregation results in a DLWA of 1 without affecting performance irrespective of device utilization. At higher utilizations, FDP improves p99 read and write latency.} 
        \label{fig:waf:kvcache-varying-util}
    \end{minipage} \hfill    
\end{figure*}
Figure \ref{fig:waf:kvcache-varying-util} shows the impact of varying the SSD capacity used for caching upon various important metrics in CacheLib. This experiment quantifies the effect of increasing the host-level utilization of the SSD for caching. We see that the DLWA increases from 1.3 at 50\% utilization to 3.5 at 100\% utilization when data segregation is not employed, but the DLWA remains unchanged at \textasciitilde1.03 across utilizations when data segregation is employed. We omit presentation of data points for utilization between 50\% and 90\% for brevity because they are similar to 50\%. This result validates our analysis of device overprovisioning and the SOC invalidation rate being key factors affecting the DLWA. At 4\% SOC size, the collision rate of SOC buckets is very high. This results in a high invalidation rate of SOC data in the SSD. As SOC data is segregated from LOC data, this high invalidation rate will result in minimal valid data movement due to garbage collection. Furthermore, the SSD device overprovisioning capacity, which ranges from 7-20\% of SSD capacity, can now be used exclusively by SOC data to reduce the impact of garbage collection. Since the random writes of SOC constitute only 4\% of the SSD capacity, their impact even at 100\% device utilization is absorbed by the extra free blocks reserved by the the SSD. \textit{We can see that the throughput, DRAM and NVM cache hit ratio metrics remain unchanged with FDP-based segregation compared to the baseline. The p99 read and write latency also show improvement with increasing utilization due to lower interference from the garbage collection process.} At 100\% device utilization, the p99 read and write latency shows an improvement of 1.75X and 10X respectively. \textbf{Since we made no changes to how data is stored in SOC and LOC, we did not expect to see any change in the ALWA which we confirmed from the CacheBench logs.}

\subsection{FDP-based segregation lowers DLWA with other write-intensive workloads}
\label{sec:eval:waf:varying-workloads}
\begin{figure*}[!t]
    \centering
    \begin{minipage}{\columnwidth}
        \includegraphics[width=0.46\linewidth]{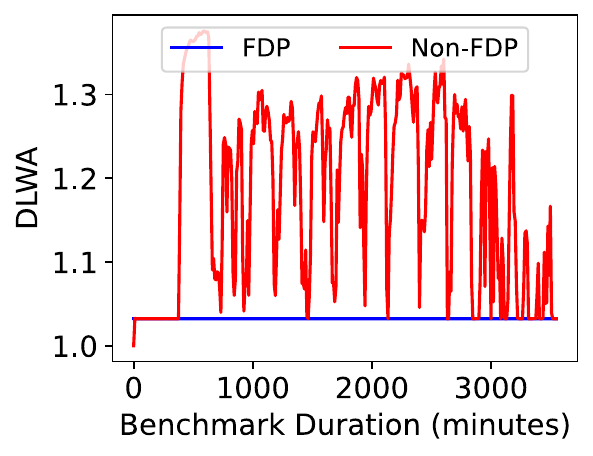}
        \hfill
        \includegraphics[width=0.46\linewidth]{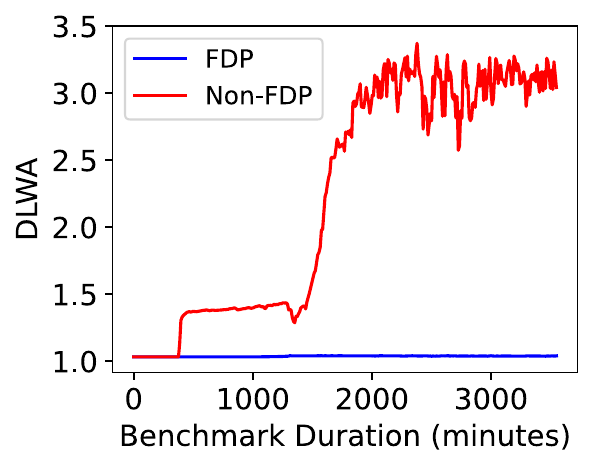}
    
        \caption{DLWA over 60 hours with the Twitter cluster 12 workloads with 16GB RAM, 4\% SOC size. a.) 50\% device utilization and b.) 100\% device utilization. FDP-based segregation achieves a DLWA of 1.}
        \label{fig:twt}
    \end{minipage}\hfill
    \begin{minipage}{\columnwidth}
    \includegraphics[width=0.46\linewidth]{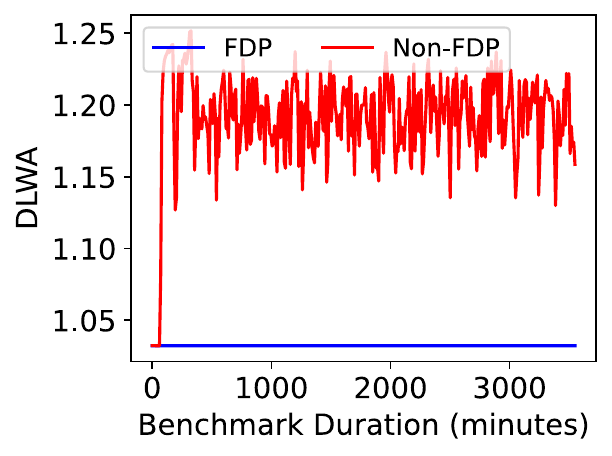}
    \hfill
    \includegraphics[width=0.46\linewidth]{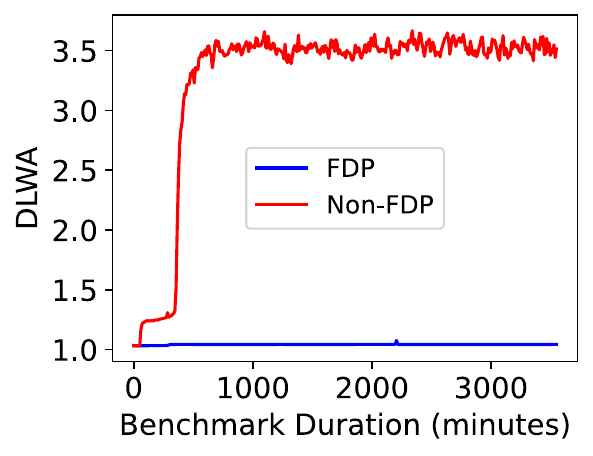}
        \caption{DLWA over 60 hours with the WO KV Cache workload with 42 GB RAM and 4\% SOC size. a.) 50\% device utilization and b.) 100\% device utilization. FDP-based segregation achieves a DLWA of 1.} \label{fig:wo-kv-waf}
    \end{minipage}
\end{figure*}
In previous sections, we demonstrated the effect of the FDP-based segregation on DLWA with the read-intensive KV Cache workload. We now explore the efficacy of our solution with the write-heavy Twitter cluster12 workload and the write-only KV Cache workload (WO KV Cache). We run experiments with both workloads at 50\% and 100\% device utilization, 4\% of SOC size and use $\sim$16GB and $\sim$42GB of DRAM for the Twitter and WO KV cache workloads respectively as used in past research~\cite{berg2020cachelib, mcallister2021kangaroo}. From Figure \ref{fig:twt} and \ref{fig:wo-kv-waf}, we can see that the DLWA trends observed with the KV Cache workload at 50\% and 100\% device utilizations remain consistent with the Twitter and WO KV Cache workloads as well. \textit{The FDP-based segregation achieves a DLWA of $\sim$1 with both these challenging write-intensive workloads dominant in small object accesses.}

\subsection{FDP-based segregation gains diminish with increase in SOC size}
\begin{figure}[!hb]
  \centering
  \includegraphics[width=0.8\linewidth]{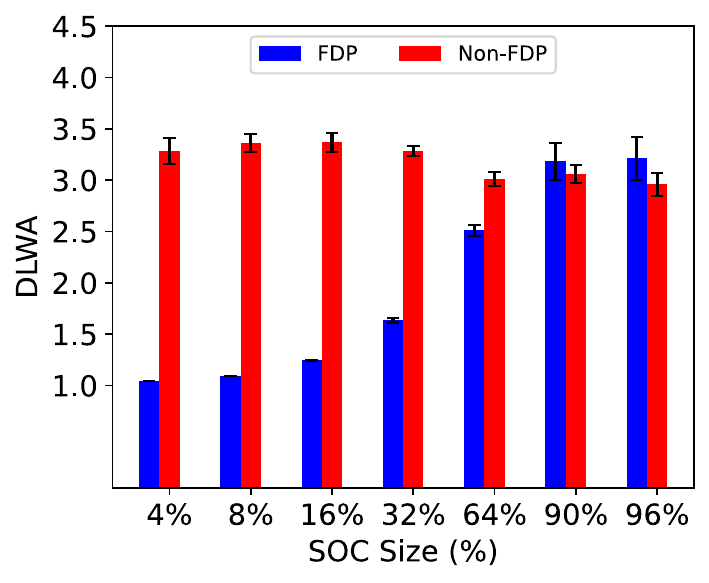}
  \caption{Average DLWA with the KV Cache workload using 100\% device utilization, 42GB of RAM and varying SOC size from 4\% to 96\% of the SSD size. FDP's DLWA gains diminish with an increase in the SOC size beyond the device overprovisioning size.} \label{fig:waf:kv-vary-soc}
\end{figure}
In previous sections, we observed excellent DLWA behaviour for the KV cache workload primarily because our implementation segregates SOC's random and hot data from LOC's sequential and cold data. The small SOC size of 4\% paired with data segregation enabled a high invalidation of SOC data in the SSD and allowed device overprovisioning to efficiently cushion garbage collection of SOC data. To further validate our analysis, we study the impact of increase of SOC size (random writes) on DLWA beyond the device overprovisioning size, which is typically 7-20\% of SSD capacity. We can see from Figure \ref{fig:waf:kv-vary-soc} that when SOC size is increased from the default 4\% to 64\% the DLWA of our implementation increases from 1.03 to 2.5 while the DLWA without segregation remains above 3 for all SOC sizes. 

As the SOC size crosses the device overprovisioning size, the DLWA with FDP no longer remains 1. The high DLWA at larger SOC sizes occurs because the spare blocks are fewer than the SOC data blocks. Consequently, the spare blocks coming from the device overprovisioning space fail to provide an adequate cushion for the garbage collection of SOC data. \edit{Despite the lack of cushioning, we can see that data segregation is helpful in invalidating the SOC data blocks and reduces movement of LOC data upon garbage collection. At very high SOC sizes e.g., 90\% and 96\% we observe that data segregation does not yield any benefits. At very high SOC sizes, there is a high probability of erase blocks containing both valid and invalid SOC data. In this scenario, data intermixing might be beneficial since sharing of invalid SOC and LOC data in an erase block would minimize data movement. Moreover, garbage collection has a lower threshold when FDP is enabled compared to when it is disabled that may accentuate the DLWA of data segregation. We observe that increasing the SOC size does not benefit the cache behavior of the workload dominant in small objects since the hit ratio almost remains unchanged.}

\subsection{FDP-based segregation enables cost-effective and carbon-efficient CacheLib deployments}
In previous sections, we showed large gains in DLWA from using FDP-based segregation in CacheLib. These gains translate to a longer SSD lifetime that leads to reductions in embodied carbon emissions. The DLWA gains also aid in improving the operational efficiency of SSDs due to fewer device garbage collection operations. This results in a reduction in operational carbon emissions. In this section, we discuss the promise of FDP as a sustainable solution to combat carbon emissions.

\begin{figure}[!ht]
  \centering
  \includegraphics[width=\linewidth]{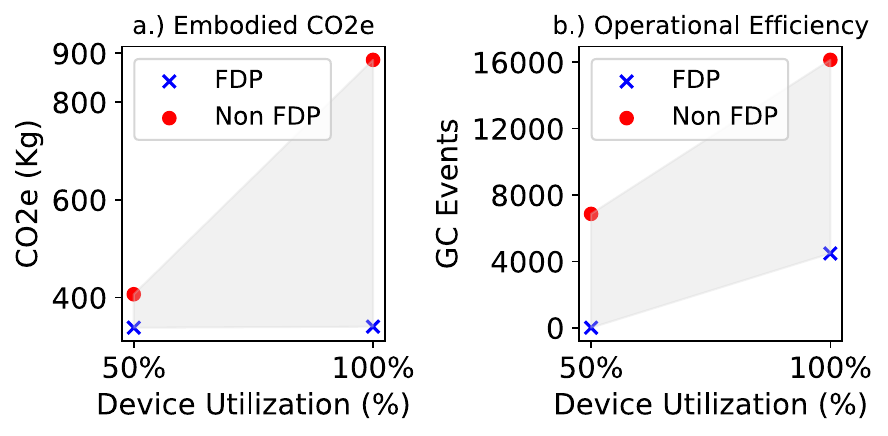}
  \caption{Analysis of carbon savings on FDP vs Non-FDP with the KV Cache workload. a.) Embodied carbon emissions reduce drastically with FDP and b.) Garbage Collection events are reduced by a factor of $\sim$3.6 with FDP.} \label{fig:kv-co2e}
\end{figure}
\minisec{FDP-based segregation reduces carbon emissions}
To calculate the embodied emissions, we use Theorem 2 presented in Section \ref{theorem:embodied-co2e} which models the embodied carbon emissions as a function of DLWA, system lifecycle period, SSD warranty, and carbon emitted by the SSD manufacturing process. We use a system lifecycle and SSD warranty of 5 years and 0.16 CO2e (Kg) as the carbon emitted per GB of SSD manufactured~\cite{embodiedcarbon}. Figure \ref{fig:kv-co2e}a.) shows the embodied carbon emissions with and without FDP-based segregation in CacheLib. We can see that FDP enables substantial embodied carbon savings as a result of the DLWA gains. These embodied carbon gains are for a single SSD over a 5-year lifecycle. If we factor the deployment of 1000s of CacheLib clusters each consisting of 1000s of nodes, the embodied carbon emission gains from using FDP are massive.

The operational energy consumption of an SSD is directly proportional to the garbage collection events (see Theorem 3 in Section \ref{theorem:embodied-co2e}). Utilizing FDP's \texttt{Media Relocated} Event in the SSD log~\cite{fdp_tp} we calculate the total number of garbage collection events that occurred when FDP is enabled. For the case of Non-FDP, we run the experiment with FDP enabled but force SOC and LOC to use a single RUH to simulate the Non-FDP scenario. We ensure the garbage collection events occur for the same amount of host writes to the SSD to account for the energy consumption due to internal operations. Figure \ref{fig:kv-co2e}b.) shows that with the KV Cache workload, FDP-based segregation resulted in $\sim$3.6x fewer GC events for the same amount of host writes to the SSD. This shows that FDP-based segregation helps in improving the operational energy consumption of the SSD leading to operational carbon savings. 
Based on DLWA gains obtained for Twitter and WO KV Cache in Section \ref{sec:eval:waf:varying-workloads}, large embodied and operational carbon gains can be realized for both those workloads. \\

\minisec{FDP-based segregation allows for carbon-efficient deployments with lower DRAM requirements}
\begin{table}[t]
    \centering
    \begin{tabular}{lcccc}
\toprule
   \makecell{Configuration \\  }& \makecell{Hit Ratio \\ (\%)} & \makecell{NVM Hit \\ Ratio (\%)} & KGET/s & \makecell{CO2e \\ (Kg)}  \\
\midrule
    FDP 4GB & 86.3 & 37.74 & 303.1 & 347.2 \\
    Non-FDP 4GB & 86.11 & 37.41 & 298.8 & 1081.1 \\
    \midrule
    FDP 20GB &  91.9  & 31.37 & 412.2 &  372.8\\
    Non-FDP 20GB &  92.1 &  33 &  399.1 & 1106.8\\
    \midrule
    FDP 42GB & 90.32 & 17.51 & 445.9 & 409.6 \\
    Non-FDP 42GB & 90.22 & 17.34 & 434.4 & 1143.6 \\
\bottomrule
    \end{tabular}\vspace{2ex}
    \caption{KV Cache workload with 100\% device utilization, 4\% SOC size and RAM size of 4GB and 42GB. We see that FDP enables carbon-efficient deployments with reduced DRAM for a tradeoff in hit ratio and throughput.}
    \label{tab:vary-ram}
    \vspace{-2ex}
\end{table}
In this section, we explore whether increase in SSD utilization in the Flash Cache of CacheLib can help reduce the DRAM Cache size. The main motivation behind this exploration is to reduce the cost and carbon footprint of CacheLib deployments. DRAM's embodied carbon footprint is at least an order of magnitude higher than an SSD~\cite{GuptaEHWL0W22}. A similar trend also exists for cost. We run the KV Cache workload with 100\% device utilization, 4\% SOC size and vary the DRAM used in the RAM Cache layer of CacheLib. 

Table \ref{tab:vary-ram} shows the overall hit ratios, NVM hit ratios, throughput and effective carbon emission with different RAM cache sizes. We see that a lower DRAM leads to a reduction in hit ratio and throughput while being more carbon-efficient. An SSD utilization of 100\% enables more items to be cached in the Flash Cache. Consequently, we see that as the RAM cache size reduces, the NVM hit ratios improve massively for both FDP and Non-FDP while the overall hit ratios and throughput drop. \textit{A deployment with lower DRAM and 100\% device utilization was not viable without FDP-based segregation due to the high DLWA of $\sim$3.5}. While the 1.5x reduction in throughput might not be acceptable to applications with strict SLAs, applications that can tolerate this decrease for a 4x gain in carbon savings might find this deployment appealing. Since CacheLib is used as a building block for services~\cite{berg2020cachelib}, the flexibility in deployment provided by FDP to achieve greater carbon efficiency may be highly desirable. 

\subsection{FDP-based segregation enables multi-tenant KV Cache deployments}
\begin{figure}[!h]
  \centering
  \includegraphics[width=0.8\linewidth]{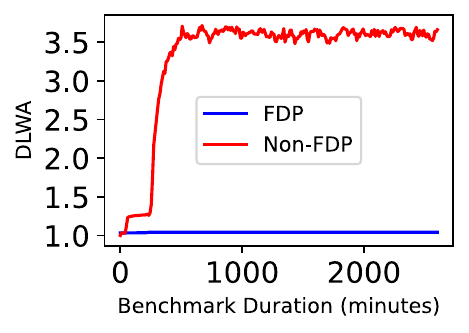}
  \caption{DLWA over 60 hours with the WO KV Cache workload running on two tenants each using 930GB SSD space and 4\% SOC size and 42 GB RAM. FDP enables a 3.5x reduction in DLWA in this multi-tenant deployment.} \label{fig:multi_tenant}
\end{figure}
Without the use of FDP, 50\% of the SSD had to be reserved for host overprovisioning to achieve an acceptable DLWA. With FDP, however, we demonstrated that a DLWA of \textasciitilde1 can be achieved without any host overprovisioning. This effectively frees up half of the device, allowing it to be utilized for other purposes. One option we explored was to increase the SSD capacity for a single CacheLib instance from 50\% to 100\%. Another option involves running two CacheLib instances that share the SSD, each using half of the available space for its Flash cache to simulate a multi-tenant setup.

We evaluate the effectiveness of SOC and LOC data segregation under this multi-tenant configuration, with two KV cache instances running the WO KV cache workload on a shared 1.88 TB SSD without any overprovisioning. We partition the SSD into two equal parts (\textasciitilde930 GB each), with each KV cache instance (tenant) assigned to one partition. Both SOC and LOC sizes are set at 4\% and 96\% respectively. Our placement policy maps the SOC and LOC of the two tenants to different reclaim unit handles while all other parameters remain unchanged. Figure \ref{fig:multi_tenant} shows that the DLWA remains \textasciitilde1 because each tenant segregates its SOC and LOC data. In contrast, without FDP the DLWA increases to \textasciitilde3.5.
\section{Related Work}
\label{sec:related-work}
\subsection{Data Placement in SSDs}
Write amplification in Flash-based SSDs~\cite{BoboilaD10:write-endurance,HeKAA17:ssd-contract,LeeSHC15:f2fs,LuSZ13:fs-reduce-write-amp,JungK13:ssd-expectations} is a well studied problem that has received attention through various data placement proposals over the past years. To tackle device-level write amplification (DLWA), SSD controllers use various heuristics to segregate data based on its characteristics (e.g., access patterns, temperature, etc.) on Flash media to minimize garbage collection and data movement costs~\cite{GalT05:ssd-survey, ChangKL04:garbage-collection}. There have been various proposals in the NVMe data placement space for cooperation between host and FTL to leverage host application domain knowledge. One approach is to pass hints to the FTL as proposed in Multi-Streamed SSDs~\cite{kang2014multi}. Another contrary approach proposed in Open-Channel SSDs~\cite{bjorling2017lightnvm} and DFS~\cite{JosephsonBLF10:dfs} is to expose the SSD internals completely paving the way for host-based FTLs. There have also been software-defined Flash proposals to expose NAND channels in SDF~\cite{OuyangLSHWW14:sdf} and to explore alternative storage abstractions that can be leveraged by key-value stores~\cite{LeeLJXKA16:app-managed-flash}. 

Zoned Namespaces (ZNS)~\cite{bjorlingAHRMGA21} was a follow-up on Open-Channel SSD 2.0 specification designed to leverage existing support of SMR HDDs. Despite impressive DLWA results, the append-only write model and host-based garbage collection imposes upfront software engineering costs for applications that do not conform to log-structured access patterns. This has posed a challenge for the wide adoption of ZNS. FDP TP~\cite{fdp_tp} was proposed based on lessons learned from the past. It consolidates Google's SmartFTL~\cite{smartftl} and Meta's Direct Placement Mode proposals to fill the cost-benefit gap between ZNS and conventional SSDs. The concept of RUHs in FDP borrows heavily from the concept of streams in multi-streamed SSDs. FDP was crafted with backward compatibility in mind, enabling an application storage stack to operate seamlessly without modifications. Additionally, applications have the option to harness FDP features by opting to enable them. FDP also does not introduce any new command sets. In this work, we have leveraged FDP features for data placement using Linux kernel I/O Passthru features~\cite{io-passthru, JoshiG0KRGLA24:iopassthru}. The backward compatibility of FDP and its ease of integration into the CacheLib software have been key drivers in the upstreaming process of our work.

\subsection{Key-Value Stores and Hybrid Caching}
A lot of research effort has been dedicated to the design of key-value stores~\cite{rocksdb, LimFAK11:silt, RajuKCA17:pebblesdb} that conform to the performance tradeoffs of Flash-based SSDs~\cite{HeKAA17:ssd-contract, JungK13:ssd-expectations,ssd_agarwal}. Their inadequacy for hybrid caching use cases at Meta that are dominated by random writes of small objects causing severe DLWA has been previously highlighted~\cite{mcallister2021kangaroo, berg2020cachelib}. Log-structured caches~\cite{ShenCJS17:didacache, EisenmanCPHSAK19:flashield} designed for Flash to minimize write amplification suffer from DRAM overhead issues (especially for numerous small items) which formed the motivation for the work in Kangaroo~\cite{mcallister2021kangaroo}. Our present work is complementary to these efforts because we keep the cache architecture and design of CacheLib~\cite{berg2020cachelib} unchanged and leverage FDP features for data placement in its I/O write paths to minimize DLWA. \edit{FairyWren~\cite{fairywren} extended Kangaroo to integrate ZNS devices for lower DLWA. We observe similar DLWA and carbon emission gains through data placement alone, without modifying the original architecture and design of CacheLib.} 
\section{Conclusion}
The problem of device-level write amplification (DLWA) is becoming exceedingly important given the increased deployment of Flash caches and the sustainability challenges faced by modern data centers. Our work of segregating Flash cache data in Flash media using FDP SSDs shows an ideal DLWA of \textasciitilde1 is possible without any host overprovisioning and overhead to Flash cache metrics. This results in massive cost (2x) and carbon emission (4x) reductions at scale. The reduction in DLWA, along with increased device utilization, opens up new deployment opportunities for hybrid caches that were previously unfeasible. For example, this approach can reduce DRAM consumption while enabling SSD sharing in multi-tenant settings. Our work to isolate data using FDP in a state-of-the-art Flash cache, CacheLib has already been merged in the upstream repository. This highlights its impact and the appeal of reduced engineering cost of FDP SSDs for Flash caches where data placement is key to reduce their DLWA and carbon emissions. 

\begin{acks}
\edit{
We would like to thank the anonymous reviewers and our shepherd, Jian Huang, for their constructive feedback and insightful comments, which greatly shaped this work. We also extend our gratitude to Samsung Memory Research Center (SMRC) for providing the infrastructure that enabled experimentation for this study, as well as for their support. Additionally, we would like to acknowledge the efforts of Ross Stenfort, Jaesoo Lee, and the entire CacheLib team at Meta for their invaluable feedback and guidance. Lastly, we are grateful to the members of Samsung’s Global Open-ecoSystem Team (GOST) for their feedback, guidance, and support in enabling the FDP ecosystem, without which this work would not have been possible.
}
\end{acks}

\bibliographystyle{ACM-Reference-Format}
\bibliography{cachelib}

\appendix
\newpage

\section{Theoretical Model of DLWA in FDP-enabled CacheLib}
\subsection{System Model: Assumptions and Observations}
\label{ref:appendix:theoretical-model}
We make the following key observations and assumptions about CacheLib's Flash cache that form the basis of our theoretical model.
\begin{itemize}[noitemsep, topsep=1pt, partopsep=0pt,leftmargin=*]
    \item Items are inserted into the LOC in a log-structured manner where evictions occur using a FIFO policy. This results in a purely sequential write pattern of LBAs to the SSD. 
    \item With FDP, the LOC writes are written into a separate RU and physical block. The purely sequential write pattern of LOC and its segregation into a separate physical block and RU means that LOC data invalidates itself nicely and contributes to negligible (or no) live data movement. 
    \item Items are inserted into the SOC buckets using a uniform hash function. Every SOC item insertion results in the entire hashed bucket (a 4KB page) being written. While a modulo is not purely sequential, for this model we assume that the SOC generates a uniform random write pattern. 
    \item With FDP, all SOC data is segregated into a separate RU and physical block. The random nature of SOC writes means that its data contributes to live data movement.
    \item As the LOC writes are sequential and will not need any live data movement, it is safe to assume that only SOC writes contribute to device-level write amplification.
    \item The SSD has some amount of overprovisioning space.
    \item In the case of FDP, since the physical blocks and RUs that contain LOC data free themselves up due to the sequential write pattern, we assume that the overprovisioned space can be entirely used by the writes from SOC data.
\end{itemize}

\begin{table}[!t]
    \centering    
    \begin{tabular}{ll}
    \toprule
        \textbf{Variable} & \textbf{Description} \\
        \midrule
        $\text{N}_{\text{B}}$ & Number of SOC Buckets \\
        $\text{N}_{\text{BB}}$  & Number of SOC Buckets per Erase Block \\
        $\text{S}_{\text{SOC}}$ & Total SOC logical space  \\
        $\text{S}_{\text{Bucket}}$ & SOC bucket size \\
        $\text{S}_{\text{Total}}$ & Total Physical Space \\
        $\text{S}_{\text{Usable}}$ & Total Usable Physical Space \\
        $\text{S}_{\text{OP}}$ & Total OP Space \\
        $\text{S}_{\text{LOC}}$ & Total LOC logical Space \\
        $\text{S}_{\text{P-LOC}}$ & Total Physical Space for LOC \\
        $\text{S}_{\text{P-SOC}}$ & Total Physical Space for SOC data \\
        $\text{S}_{\text{NVM}}$ & Total logical space used for the NVM Cache \\
        $\delta$ & Avg. live SOC bucket migrations due to GC \\
        \bottomrule
    \end{tabular}
    \caption{System Variables}
    \label{tab:params}
\end{table}

\subsection{Derivation}
We follow the same methodology as proposed in ~\cite{DayanBB15} to model the DLWA of SOC writes in the SSD. The SOC DLWA equates to the overall CacheLib DLWA because the write amplification of LOC data after data segregation with FDP is $\sim$1.

The total number of SOC buckets can be calculated as:
\begin{equation}
   \text{N}_{\text{B}} =   \frac{\text{S}_{\text{SOC}}}{\text{S}_{\text{Bucket}}}
\end{equation}
Items are inserted into buckets using a \% $\text{N}_{\text{B}}$. As the bucket size increases, the total number of SOC buckets reduces. \\

The total physical space in the SSD can be given by:
\begin{equation}  \label{eq:total_space}
    \text{S}_{\text{Total}} = \text{S}_{\text{Usable}} + \text{S}_{\text{OP}}
\end{equation}

We assume that LOC data will use $\sim$0 OP space, i.e.
\begin{equation} \label{eq:loc_assumption}
   \text{S}_{\text{LOC}} = \text{S}_{\text{P-LOC}}
\end{equation} 

Using Equation \ref{eq:total_space} and Equation \ref{eq:loc_assumption}, and assuming the entire usable space will make up the NVM Cache capacity (i.e. no host overprovisioning),

\begin{equation} \label{eq:soc_assumption}
    \text{S}_{\text{P-SOC}} = \text{S}_{\text{SOC}} + \text{S}_{\text{OP}}
\end{equation}

The LBA space of the SOC data i.e. $\text{S}_{\text{SOC}}$ is uniformly distributed. We assume that the number of SOC buckets in an erase block is $\text{N}_{\text{BB}}$. Then, it follows that after X SOC insertions the probability that a particular SOC bucket gets updated in an erase block is:
\begin{equation} \label{eq:prob_soc_re_write}
    \text{p(SOC bucket rewrite)} = \frac{\text{N}_{\text{BB}}}{\text{S}_{\text{SOC}}}
\end{equation}

This follows a geometric distribution. This is similar to other DLWA modelling work done before~\cite{DayanBB15}. 
In a geometric distribution, after X trials we get a success of a particular bucket getting overwritten or updated. In a geometric distribution, the mean is $\frac{1}{p}$. So,

\begin{equation}\label{eq:geo_mean}
    \mu = \frac{\text{S}_{\text{SOC}}}{\text{N}_{\text{BB}}}
\end{equation}

Equation \ref{eq:geo_mean} is analogous to saying that on average after $\frac{\text{S}_{\text{SOC}}}{\text{N}_{\text{BB}}}$ updates/SOC writes, the first SOC bucket gets overwritten in a particular erase block. Similarly, the second SOC bucket gets overwritten after  $\frac{\text{S}_{\text{SOC}}}{\text{N}_{\text{BB}} - 1}$ SOC inserts on average, and so on. \\ It follows that the number of updates after which there are no SOC buckets yet to be overwritten is,

\begin{equation}\label{eq:euler}
    \frac{\text{S}_{\text{SOC}}}{\text{N}_{\text{BB}}} + \frac{\text{S}_{\text{SOC}}}{\text{N}_{\text{BB}} - 1} + \frac{\text{S}_{\text{SOC}}}{\text{N}_{\text{BB}} -2} + ... + \frac{\text{S}_{\text{SOC}}}{1}
\end{equation}

This harmonic series can be simplified using Euler's approximation. 
\begin{equation}\label{eq:euler_simple}
    \text{X} =  \text{S}_{\text{SOC}} \sum_{i = 1}^{\text{N}_{\text{BB}}} \frac{1}{i}
\end{equation}

If $\text{L}$ SOC buckets are yet to be overwritten on average in an erase block after X SOC insertions. Then,
\begin{equation}\label{eq:alive_soc_buckets}
     \text{X} = \text{S}_{\text{SOC}} \sum_{i = 1}^{\text{N}_{\text{BB}}} \frac{1}{i} - \text{S}_{\text{SOC}} \sum_{i = 1}^{L} \frac{1}{i}
\end{equation}

This can be expressed as:
\begin{equation}\label{eq:alive_soc_buckets_simple}
    \text{L} = \text{N}_{\text{BB}} e^{- X/\text{S}_{\text{SOC}}}
\end{equation}

We now factor in the overprovisioned space available to the SOC data during SSD garbage collection operations to cushion the DLWA. \\

Say the average SOC buckets that remain valid in an erase block when GC is triggered is $\delta$. Then each GC operation on average involves $\text{N}_{\text{BB}} \times \delta$ migrations. It also follows that the number of SOC bucket writes or insertions that can be accommodated due to this migration is $\text{N}_{\text{BB}} \times (1- \delta) $. \\
We assume a greedy GC policy (i.e. the erase block with least valid pages will be picked first for GC). Say the expected number of GC operations between two successive victim selections of the same block is $\text{G}$. Given the uniform workload pattern we can see that, 
\begin{equation} \label{eq:pba}
    \text{G} = \frac{\text{S}_{\text{P-SOC}}}{\text{N}_{\text{BB}}}
\end{equation}
This can be interpreted as follows, 
\begin{itemize}
    \item We have $\frac{\text{S}_{\text{P-SOC}}}{\text{N}_{\text{BB}}}$ number of erase blocks holding SOC data in the SSD.
    \item Once an erase block is picked, on average it is picked after $\frac{\text{S}_{\text{P-SOC}}}{\text{N}_{\text{BB}}}$ GC operations.
\end{itemize}

As $\text{G}$ erase blocks are picked for GC before the same block is picked again, $ \text{G} \times \text{N}_{\text{BB}} \times (1- \delta) $ SOC writes occur on average between two GC operations on the same erase block. \\

Using this in Equation \ref{eq:alive_soc_buckets_simple} and Equation \ref{eq:pba} we get,

\begin{equation}\label{eq:lba_pba}
    \frac{\text{S}_{\text{SOC}}}{\text{S}_{\text{P-SOC}}} = \frac{\delta - 1}{\text{ln}(\delta)}
\end{equation}

Equation \ref{eq:lba_pba} can be simplified using the Lambert W function ~\cite{DayanBB15, Desnoyers12, StoicaA13} as follows:

\begin{equation}\label{eq:final}
    \delta = - \frac{\text{S}_{\text{SOC}}}{\text{S}_{\text{P-SOC}}} \times \mathcal{W} (- \frac{\text{S}_{\text{P-SOC}}}{\text{S}_{\text{SOC}}} \times e^{- \frac{\text{S}_{\text{P-SOC}}}{\text{S}_{\text{SOC}}}})
\end{equation}

The SSD DLWA can be calculated using $\delta$ as,
\begin{equation}\label{eq:WAF}
    DLWA = \frac{1}{1 - \delta}
\end{equation}

\subsection{Validation of the DLWA Model with FDP-enabled CacheLib Empirical Result}
\begin{figure}[!t]
  \centering
  \includegraphics[width=0.70\linewidth]{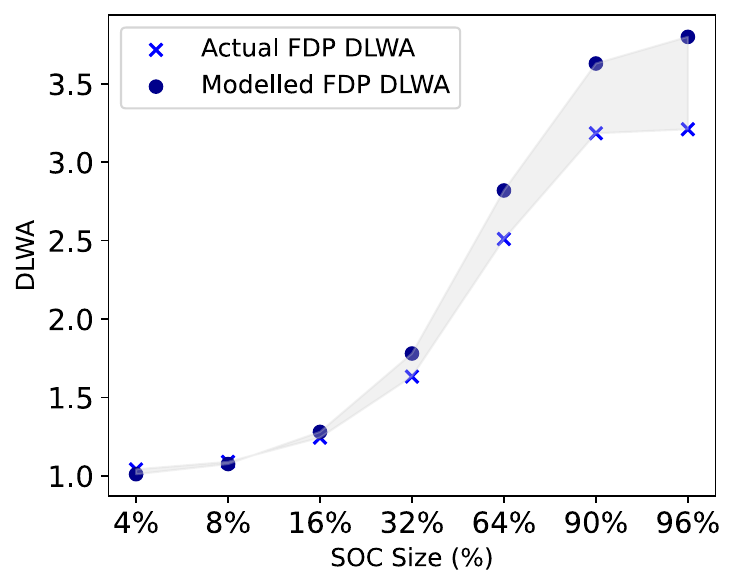}
  \caption{DLWA obtained from experiments using the KV Cache workload at 100\% device utilization, 42GB RAM and varying SOC size in comparison to the DLWA obtained from the formula. We see minimal error in DLWA estimation using the formula.} 
  \label{fig:modelling}
\end{figure}

Figure \ref{fig:modelling} shows the minimal error in estimating CacheLib's DLWA using the DLWA model (Section \ref{theorem:DLWA}) in comparison to the observed DLWA obtained from experiments with the KV Cache workload. \edit{We see that at high SOC values the model diverges by a maximum of \textasciitilde16\% from the observed DLWA. We observe that at high SOC values, the model diverges from the empirical results because it assumes a uniform distribution of keys to SOC buckets. This is actually not the case due to skew causing the observed DLWA to be lower than the predicted DLWA. Note that we ran this experiment at 100\% device utilization. At lower device utilization, the error will be similar to low SOC values due to low DLWA.}

\section{FDP-based Segregation Benefits with WO KV Cache Workload}
\begin{figure*}[!t]
        \begin{minipage}{0.95\linewidth}
        \hspace{-1em}
        \centerline{\includegraphics[width=0.95\linewidth]{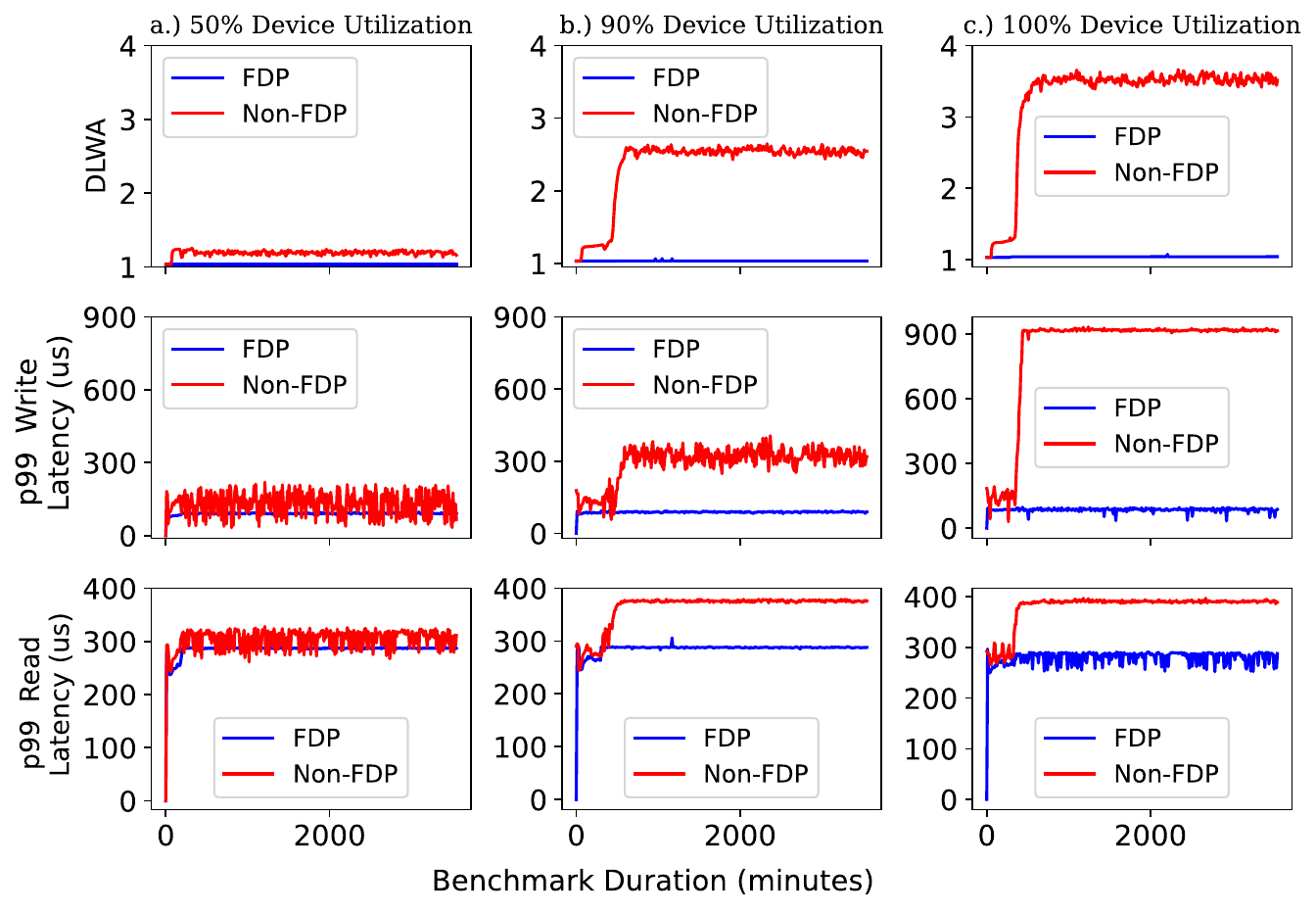}}
        \caption{Effect of varying SSD utilization for caching with WO KV Cache Workload on DLWA and p99 read and write latency. FDP-based segregation results in a DLWA of 1 without affecting performance irrespective of device utilization. At higher utilizations, FDP improves p99 read and write latency.} 
        \label{fig:waf:wo-kvcache-varying-util}
    \end{minipage}
\end{figure*}

Figure \ref{fig:waf:wo-kvcache-varying-util} shows the observed DLWA, p99 read and write latencies with FDP-enabled CacheLib using the WO KV Cache workload across different device utilization. We see similar trends as observed before with other workloads  in Section~\ref{sec:eval:waf:kv-varying-util} and Section~\ref{sec:eval:waf:varying-workloads} i.e., increasing gains in DLWA and lowering of p99 read and write latency at higher device utilizations. At 100\% device utilization, FDP-based data segregation obtains 3.5x gains in DLWA, 2.2x gains in p99 read latency, and 9.5x gains in p99 write latency.

\balance
\end{document}